\documentclass[conference]{IEEEtran}
\IEEEoverridecommandlockouts
\usepackage{amsmath,amssymb,amsfonts}
\usepackage{graphicx}
\usepackage{textcomp}
\usepackage{xcolor}
\def\BibTeX{{\rm B\kern-.05em{\sc i\kern-.025em b}\kern-.08em
    T\kern-.1667em\lower.7ex\hbox{E}\kern-.125emX}}

\usepackage[english]{babel}
\usepackage{csquotes}
\usepackage{amsmath}
\usepackage{mathtools}

\usepackage[backend=bibtex]{biblatex}
\usepackage{graphicx}
\usepackage{import}
\usepackage{color}
\usepackage[hidelinks]{hyperref}
\usepackage{url}
\usepackage{tikz}
\usetikzlibrary{patterns}
\usepackage{pgfplots}
\usepackage[binary-units=true]{siunitx}
\usepackage{booktabs}
\usepackage{multirow}
\usepackage{pifont}
\usepackage{eurosym}
\usepackage{graphicx} 
\usepackage{algorithm}
\usepackage{algpseudocode}
\usepackage{caption}
\usepackage{subcaption}
\usepackage{csquotes}
\usepackage[capitalize]{cleveref}
\usepackage[acronym, nonumberlist, toc, nopostdot]{glossaries} %
\definecolor{tumblue}{RGB}{0,101,189}
\definecolor{darkgreen}{RGB}{55,157,47}

\usepackage[binary-units=true]{siunitx}

\definecolor{myRed}{RGB}{165,15,21}
\definecolor{myBlue}{RGB}{49,130,189}
\definecolor{myOrange}{RGB}{253,141,60}
\definecolor{myGreen}{RGB}{237,248,233}

\makeatletter
\renewcommand*\env@matrix[1][\arraystretch]{%
  \edef\arraystretch{#1}%
  \hskip -\arraycolsep
  \let\@ifnextchar\new@ifnextchar
  \array{*\c@MaxMatrixCols c}}
\makeatother

\newif\ifnotes\notesfalse
\ifnotes
	\usepackage[colorinlistoftodos,prependcaption]{todonotes}
	\usepackage[absolute]{textpos}
	\newcommand{\TODO}[1]{\todo[inline,linecolor=black,backgroundcolor=gray!10,bordercolor=red]{#1}\noindent}
	\newcommand{\CORRECTION}[1]{\todo[inline,linecolor=red,backgroundcolor=gray!10,bordercolor=black]{#1}\noindent}
	\newcommand{\noteswarning}{\begin{textblock}{200}(10,5) \begin{center} \textcolor{red}{\LARGE 	\textbf{WARNING: NOTES ARE ON}} \end{center} \end{textblock}
	\begin{textblock}{200}(10,12.5) \begin{center}  \texttt{\tiny \IfFileExists{./revision}{\input{./revision}}{}} \end{center} \end{textblock}}
\else
	\newcommand{\TODO}[1]{}
	\newcommand{\CORRECTION}[1]{}
	\newcommand{\noteswarning}{}
\fi

\newcommand{\mytitle}{Statistical Ineffective Fault Analysis of \\\textsc{Gimli}}

\newif\ifblinded\blindedfalse

\ifblinded
\title{\mytitle\\
}

\author{\IEEEauthorblockN{Blinded for review}
	\IEEEauthorblockA{\textit{University} \\
		\textit{Institute}\\
		City, Country\\
		Email}
}
	\newcommand{\ack}[1]{Blinded for review}
\else
\title{\mytitle\\
}
\author{\IEEEauthorblockN{Michael Gruber, Matthias Probst, Michael Tempelmeier}
	\IEEEauthorblockA{\textit{Chair of Security in Information Technology} \\
		\textit{Technical University of Munich}\\
		Munich, Germany \\
		\{m.gruber, matthias.probst, michael.tempelmeier\}@tum.de}
}
	\newcommand{\ack}[1]{#1}
\fi

\definecolor{tumblue}{RGB}{0,101,189}
\definecolor{darkgreen}{RGB}{55,157,47}
\definecolor{darkred}{RGB}{197,47,47}
\definecolor{tumorange}{RGB}{207,094,024}

 \makeatletter
\newcommand*{\pmlll}{%
  \mathrel{%
    \mathpalette\@lllggg<%
  }%
}
\newcommand*{\pmggg}{%
  \mathrel{%
    \mathpalette\@lllggg>%
  }%
}
\newcommand*{\@lllggg}[2]{%
  \sbox0{$\m@th#1#2$}%
  \copy0 %
  \kern-.6\wd0 %
  \copy0 %
  \kern-.6\wd0 %
  \copy0 %
}
\makeatother
\newacronym{SCA}{SCA}{Side Channel Analysis}
\newacronym{SEI}{SEI}{Squared Euclidean Imbalance}
\newacronym{FDT}{FDT}{Fault Distribution Table}

\newacronym{SIFA}{SIFA}{Statistical Ineffective Fault Analysis}
\newacronym{SFA}{SFA}{Statistical Fault Analysis}
\newacronym{IFA}{IFA}{Ineffective Fault Analysis}
\newacronym{DFA}{DFA}{Differential Fault Analysis}
\newacronym{DFIA}{DFIA}{Differential Fault Intensity Analysis}

\newacronym{DPA}{DPA}{Differential Power Analysis}

\newacronym{AES}{AES}{Advanced Encryption Standard}

\newacronym{NIST}{NIST}{National Institute of Standards and Technology}
\newacronym{CSRC}{CSRC}{Computer Security Resource Center}
\newacronym{AD}{AD}{Associated Data}
\newacronym{AE}{AE}{Authenticated Encryption}
\newacronym{AEAD}{AEAD}{Authenticated Encryption with Associated Data}
\newacronym{HW}{HW}{hardware}
\newacronym{EM}{EM}{electro-magnetic}

\newacronym{SPBOX}{SP-Box}{Substitution Permutation Box}

\newglossary[slg]{symbols}{syi}{syg}{List of symbols} %
\makeglossaries %
\renewcommand{\arraystretch}{1.5} %
\renewcommand{\thetable}{\arabic{table}} %
\begin{document}

\thispagestyle{empty}
\onecolumn

\noindent\rule{\textwidth}{1pt}

\begin{center}
\textbf{\huge IEEE Copyright Notice}
\end{center}

\vspace{1cm}

\noindent\copyright~2019 IEEE. Personal use of this material is permitted. Permission from IEEE must be obtained for all other uses, in any current or future media, including reprinting/republishing this material for advertising or promotional purposes, creating new collective works, for resale or redistribution to servers or lists, or reuse of any copyrighted component of this work in other works.

\vspace{1cm}

\noindent Accepted to be Published in:\\
Proceedings of the \textbf{2020 IEEE International Symposium on Hardware Oriented Security and Trust (HOST)}\\ May 4-7, 2020, San Jose, CA, USA

\vspace{1cm}

\noindent This version of the paper is the one which was accepted during the review process.

\vspace{1cm}

\noindent The DOI will be added after publication.\\
\noindent\rule{\textwidth}{1pt}

\twocolumn
 
\noteswarning

\interfootnotelinepenalty=10000
\IEEEtitleabstractindextext{%
\begin{abstract}
Statistical Ineffective Fault Analysis (SIFA) was introduced as a new approach to attack block ciphers at CHES 2018.
Since then, they have been proven to be a powerful class of attacks, with an easy to achieve fault model.
One of the main benefits of SIFA is to overcome detection-based and infection-based countermeasures.
In this paper we explain how the principles of SIFA can be applied to \textsc{Gimli}, an authenticated encryption cipher participating the NIST-LWC competition.
We identified two possible rounds during the intialization phase of \textsc{Gimli} to mount our attack.
If we attack the first location we are able to recover 3 bits of the key uniquely and the parity of 8 key-bits organized in 3 sums using 180 ineffective faults per biased single intermediate bit.
If we attack the second location we are able to recover 15 bits of the key uniquely and the parity of 22 key-bits organized in 7 sums using 340 ineffective faults per biased intermediate bit.
Furthermore, we investigated the influence of the fault model on the rate of ineffective faults in \textsc{Gimli}.
Finally, we verify the efficiency of our attacks by means of simulation.\\
\end{abstract}

}
 \maketitle
\IEEEdisplaynontitleabstractindextext

\section{Introduction}
\noindent Fault attacks span a class of implementation level attack which were first introduced by Boneh et al. in their seminal work~\cite{Boneh2000}.

\noindent Fault attacks pose a serious threat to the implementation of cryptographic algorithms.
They usually aim for  a modification of the processed data or the control flow in order to reduce the underlying mathematical problem to a simpler one.
\noindent The most common type of fault attack is the Differential Fault Analysis (DFA) which requires knowledge of multiple faulted encryptions and a correct one.
\gls{SIFA}, as introduced by Dobraunig et~al.~\cite{Dobraunig2018sifa}, requires only an intermediate state with a biased distribution i.e. a distribution which deviates from the uniform distribution.
Also, in contrast to \gls{DFA} it is not necessary to have tuples of faulty and correct encryptions.
Even worse, SIFA can break traditional countermeasures against fault attacks like detection-based or infection-based countermeasures.
\\\noindent \textbf{State of the art:}
So far the priciples of \gls{SIFA} were applied to a variety of cryptographic algorithms ranging from block ciphers as AES \cite{Dobraunig2018sifa} to authentificated encryption schemes like \textsc{Ketje}, \textsc{Keyak} \cite{Dobraunig2016fault} and \textsc{Ascon} \cite{Ramezanpour2019a}.
\\\noindent \textbf{Contributions:}
We present the first \gls{SIFA} on the NIST-LWC-candidate \textsc{Gimli}.
Additionally, we verify the efficiency of our attacks by means of simulation.
In addition, we evaluate the influence of the fault model on the rate of ineffective faults.
\\\noindent \textbf{Organization:}
The rest of the work is structured as follows:
\cref{sec:lightweight} introduces authenticated encryption and the \mbox{NIST-LWC} competition.
\cref{sec:sifa} explains the principles of SIFA.
\cref{sec:gimli} presents the \mbox{NIST-LWC} candidate \textsc{Gimli}.
\cref{sec:sifa_gimli} introduces our \gls{SIFA} on \textsc{Gimli}.
\cref{sec:results} provides the results of our proposed attack on \textsc{Gimli}.%
\section{NIST Lightweight Cryptography}\label{sec:lightweight}
\begin{figure}
	\centering
	\includegraphics[width= 0.3 \textwidth]{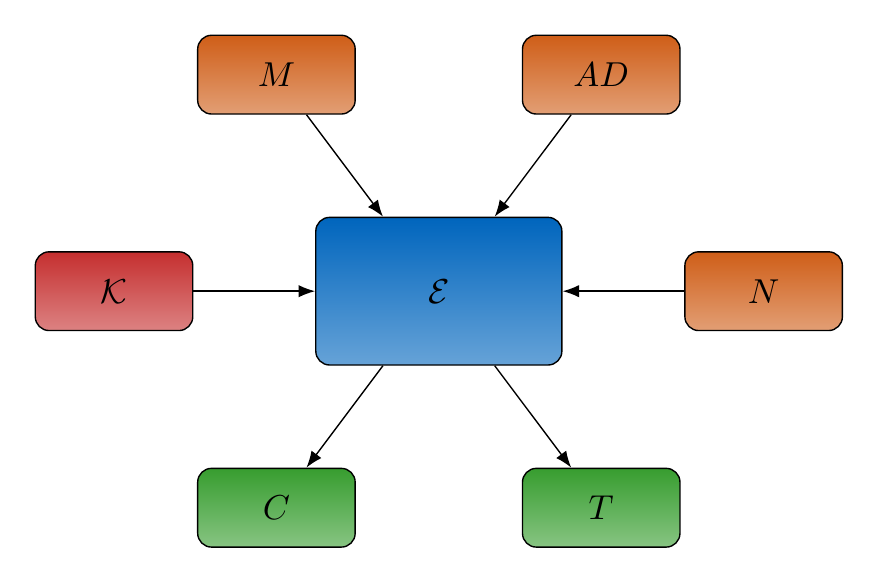}
	\caption{AEAD Encryption}
	\label{fig:lightweight_AEAD}
\end{figure}

\noindent With the growing amount of interconnected devices, the need for secure communications is ubiquitous.
While today's cryptographic algorithms are well suited for high-end computing, like servers, or desktops,
their performance degenerates dramatically, when implemented on small, resource-constrained devices.
Therefore, the \gls{NIST} launched an open standardization project for lightweight cryptography in~2013~\cite{mckay2016report}.
In~2019, the tremendous amount of 57 round~1 submissions confirms the need and interest of lightweight cryptography.
\gls{NIST} focuses on small, but secure algorithms that provide both \gls{AEAD} and Hash.
Additional features like post-quantum resistance or ease of side-channel and fault-attack resistant implementations are desirable, but not mandatory. 
In this paper, we will focus on \gls{AEAD}:
\gls{AE} combines the traditionally separated cryptographic primitives \emph{symmetric key cryptography} and \emph{authentication} into one single algorithm.
Additionally, in \gls{AEAD}, \gls{AD} is authenticated, but not encrypted.
This is particular useful for header information in a communication protocol that must be authenticated, but do not need to be encrypted.
If the authentication of either the \gls{AD} or message fails, an empty string is output, otherwise the plaintext is released. This prevents chosen ciphertext attacks.
\cref{fig:lightweight_AEAD} shows the inputs and outputs of an \gls{AEAD} scheme (encryption).
Formally, let $k,\nu,t\geq 1$, $K\in\{0,1\}^k$ denote a
secret key, $N\in\{0,1\}^\nu$ a nonce, $AD\in\{0,1\}^*$ associated data, $M\in\{0,1\}^*$ a message, 
$T\in\{0,1\}^t$ an authentication tag, and $C\in\{0,1\}^*$ a ciphertext. 
An \gls{AEAD} is a triple $\Pi = (\mathcal{K},\mathcal{E},\mathcal{D})$, with a key-generation procedure 
$\mathcal{K}$ that returns a random $K$, an encryption algorithm
$\mathcal{E}_K(N,AD,M)$, and a decryption algorithm $\mathcal{D}_K(N,AD,C,T)$, 
where $\mathcal{E}$ outputs a pair $(C,T)$, and $\mathcal{D}$ outputs either the plaintext $M$ or 
the void symbol $\perp$ if the tag is invalid:
{\small
\[
\mathcal{E}: \{0,1\}^k \times \{0,1\}^\nu \times \{0,1\}^* \times \{0,1\}^* \to \{0,1\}^*\times \{0,1\}^t
\]
\[
\mathcal{D}: \hspace{-0.5mm} \{0,1\}^k \times \{0,1\}^\nu \times \{0,1\}^* \times \{0,1\}^* \times \{0,1\}^t \hspace{-1mm} \to  \{0,1\}^*\cup \{\perp\}
\]}
\section{SIFA}\label{sec:sifa}
\noindent The exploitation of statistical ineffective faults was first proposed by Dobraunig~et~al., in order to break symmetric cryptography \cite{Dobraunig2018sifa}.
\gls{SIFA} can be seen as the combination of \gls{IFA}, as proposed by Clavier and Christophe \cite{Clavier2007secret}, and \gls{SFA} as proposed by Fuhr et al. \cite{Fuhr2013fault}.
In the following, the principles of \gls{SIFA} are explained.%
\subsection{Background} \label{subsection:sifa_background_benefits}
\noindent \gls{SIFA} combines the benefits of %
\gls{IFA} and \gls{SFA} in terms of the required fault model, the key distinguisher and the ability of overcoming countermeasures:
The required fault model of \gls{IFA}, is rather specific e.g. Clavier et al. proposed a fault model where the computation of an XOR results always in a zero value \cite{Clavier2007secret}.
The downside of the assumed fault model is the fact that the required fault model is difficult to achieve in practice.
In contrast, the assumptions of the required fault model for a \gls{SFA} are loose as the only requirement for
a successful \gls{SFA} is a biased intermediate state as shown by Fuhr et al. \cite{Fuhr2013fault}. 
Both attacks differ significantly in how they recover the key: In \gls{IFA} the recovery of the correct target partial sub key is strictly analytical; whereas in \gls{SFA} a statistical approach is used. %
One of the main benefits of the statistical approach, in \gls{SFA}, is the immunity to noisy faults i.e. injections that do not comply with the required fault model.
The main benefit of ineffective faults is the ability to break common countermeasures against fault attacks.
There are two possible countermeasures against fault attacks: detection-based countermeasures and infection-based countermeasures \cite{Dobraunig2018sifa}.
The most common form of detection-based countermeasures is temporal redundancy, where an encryption or decryption is performed twice.
If the results do not match, a fault occurred and appropriate measures like a system-reset can be taken.
The infection-based countermeasure applies additional operations in order to increase the fault propagation to an level where an attack is no longer feasible \cite{Patranabis2015,Zhang2016against}.
\begin{figure}
	\begin{center}
		\includegraphics[width = 0.4\textwidth]{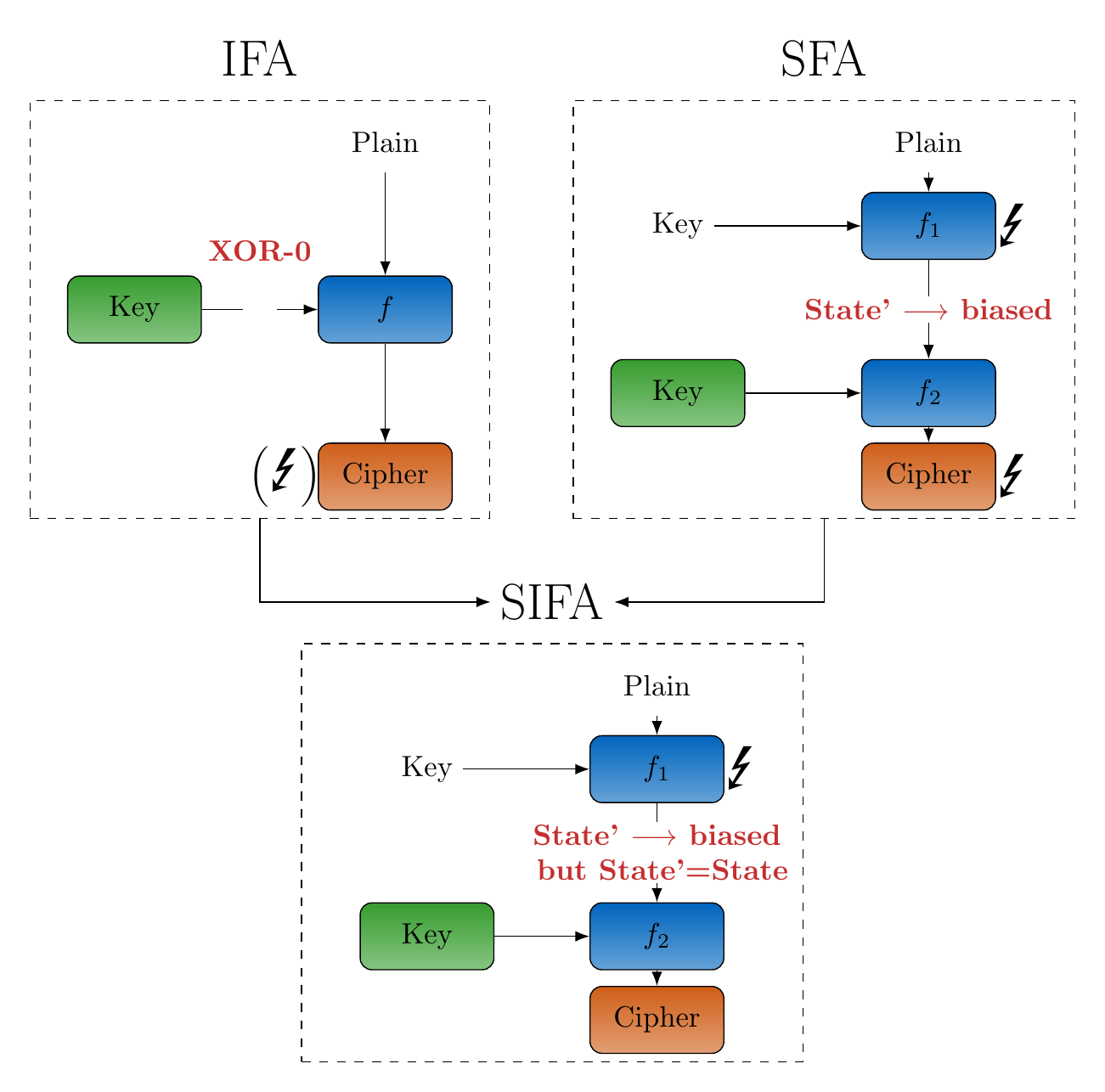}
		\caption{SIFA Background}
		\label{figure:sifa_background_benefits}
	\end{center}
\end{figure}
\gls{IFA} is based on a very precise fault model where a faulted operation always returns the same value in the general case.
By forcing the output of an operation to a specific value, the attacker can distinguish, if the output of the faulty operation is equal to the fault free output.
If the faulted output equals the fault free output, an ineffective fault occurred.
\cref{figure:sifa_background_benefits}-IFA shows this behavior, where the computation of an XOR always returns zero as in \cite{Clavier2007secret}.
\gls{IFA} is feasible even if the previously introduced countermeasures are implemented because the output of the cryptographic operation is correct.
In contrast to \gls{IFA}, \gls{SFA} exploits the non-uniform distribution of intermediate values in conjunction with the corresponding faulty output to recover the secret key.
As shown in \cref{figure:sifa_background_benefits}-SFA an attacker injects a fault after the computation of $f_1$ and before $f_2$, where $f_{i}~|~i \in \{1,2\}$ denotes parts of the cryptographic operation.
The intermediate state becomes therefore fautly and follows a biased distribution which enables a \gls{SFA}.
The attack partially decrypts the faulty ciphertext with a hypothesis of the target partial subkey in order to calculate the faulty (biased) intermediate value.
By repeating this for all key-hypotheses and ranking the corresponding intermediate values according to their deviation from the uniform distribution.
When \gls{SIFA} is applied, an attacker can determine the correct key by introducing ineffective faults.
However, in contrast to \gls{IFA} the fault model does not need to be known exactly. It is sufficient to bias an intermediate value.
Furthermore, the cipher-output needed for the attack is correct in contrast to \gls{SFA}.
\subsection{Foundations of Statistical Ineffective Fault Analysis} \label{subsection:sifa_principle}
\noindent \gls{SIFA} evaluates the statistical distribution of intermediate values and identifies appropriate key candidates with a statistical model.
\\\noindent \textbf{Distribution of Intermediate values:}
We assume that a fault only corrupts a part $s$ of the intermediate state.
The partial intermediate state after a fault injection is denoted by $s'$.
The alternation of $s \rightarrow s'~|~ s \neq s'$ results in a faulty computation and will affect the outcome of the cryptographic algorithm.
However, an alternation $s \rightarrow s'~|~s' = s$ does not affect the outcome of the cryptographic algorithm.
From now on we will refer to such a behavior as an ineffective fault.

\noindent Such ineffective faults can be exploited, if they cause a biased distribution in the intermediate value.
The six \glspl{FDT} in \cref{table:sifa_fault_distribution_table} show the transition probability of two-bit intermediate values for six typical fault models.
In order to apply \gls{SIFA}, the diagonal of such a table (marked in \textcolor{tumblue}{blue}) must differ from the uniform distribution. This holds true for \cref{stable:distribution_stuckat_0,stable:distribution_random_or,stable:distribution_random_and,stable:distribution_probabilistic_bitflip}.
The table's entries which are not colored \textcolor{tumblue}{blue} denote the probability of effective faults.
\begin{table}[ht]
	\small
	\begin{center}
		\begin{subtable}{.24\textwidth}
			\caption{Random-Or}
			\label{stable:distribution_random_or}
			\centering
			\begin{tabular}{cc|cccc}
				\multicolumn{2}{c}{} &\multicolumn{4}{c}{$s'$}\\
				&& $00$ & $01$ & $10$ & $11$ \\
				\cline{2-6}
				\multirow{4}{*}{$s$} 
				&$00$	& \textcolor{tumblue}{$\frac{1}{4}$} & $\frac{1}{4}$ & $\frac{1}{4}$ & $\frac{1}{4}$ \\
				&$01$ 	& $0$ & \textcolor{tumblue}{$\frac{1}{2}$} & $0$ & $\frac{1}{2}$ \\
				&$10$	& $0$ & $0$ & \textcolor{tumblue}{$\frac{1}{2}$} & $\frac{1}{2}$ \\
				&$11$	& $0$ & $0$ & $0$ & \textcolor{tumblue}{$1$} \\
			\end{tabular}
		\end{subtable}
		\begin{subtable}{.24\textwidth}
			\caption{Random-And}
			\label{stable:distribution_random_and}
			\centering
			\begin{tabular}{cc|cccc}
				\multicolumn{2}{c}{} &\multicolumn{4}{c}{$s'$}\\
				&& $00$ & $01$ & $10$ & $11$ \\
				\cline{2-6}
				\multirow{4}{*}{$s$} 
				&$00$	& \textcolor{tumblue}{$1$} & $0$ & $0$ & $0$ \\
				&$01$ 	& $\frac{1}{2}$ & \textcolor{tumblue}{$\frac{1}{2}$} & $0$ & $0$ \\
				&$10$	& $\frac{1}{2}$ & $0$ & \textcolor{tumblue}{$\frac{1}{2}$} & $0$ \\
				&$11$	& $\frac{1}{4}$ & $\frac{1}{4}$ & $\frac{1}{4}$ & \textcolor{tumblue}{$\frac{1}{4}$} \\
			\end{tabular}
		\end{subtable}
		\vspace{0.5cm}
		
	\begin{subtable}{.24\textwidth}
		\caption{Stuck-at-0}
		\label{stable:distribution_stuckat_0}
		\centering
	\begin{tabular}{cc|cccc}
		\multicolumn{2}{c}{} &\multicolumn{4}{c}{$s'$}\\
			&& $00$ & $01$ & $10$ & $11$ \\
		\cline{2-6}
		\multirow{4}{*}{$s$} 
		&$00$	& \textcolor{tumblue}{$1$} & $0$ & $0$ & $0$ \\
		&$01$	& $1$ & \textcolor{tumblue}{$0$} & $0$ & $0$ \\
		&$10$	& $1$ & $0$ & \textcolor{tumblue}{$0$} & $0$ \\
		&$11$	& $1$ & $0$ & $0$ & \textcolor{tumblue}{$0$} \\
	\end{tabular}
	\end{subtable}
	\begin{subtable}{.24\textwidth}
	\caption{Probabilistic bit-flip}
	\label{stable:distribution_probabilistic_bitflip}
	\centering
	\begin{tabular}{cc|cccc}
		\multicolumn{2}{c}{} &\multicolumn{4}{c}{$s'$}\\
		&& $00$ & $01$ & $10$ & $11$ \\
		\cline{2-6}
		\multirow{4}{*}{$s$} 
		&$00$	& $\textcolor{tumblue}{\frac{4}{9}}$ & $\frac{2}{9}$ & $\frac{2}{9}$ & $\frac{1}{9}$ \\
		&$01$	& $\frac{4}{9}$ & $\textcolor{tumblue}{\frac{2}{9}}$ & $\frac{2}{9}$ & $\frac{1}{9}$ \\
		&$10$	& $\frac{4}{9}$ & $\frac{2}{9}$ & $\textcolor{tumblue}{\frac{2}{9}}$ & $\frac{1}{9}$ \\
		&$11$	& $\frac{4}{9}$ & $\frac{2}{9}$ & $\frac{2}{9}$ & \textcolor{tumblue}{$\frac{1}{9}$} \\
	\end{tabular}
\end{subtable}
	
	\vspace{0.5cm}

	\begin{subtable}{.24\textwidth}
		\caption{Random fault}
		\label{stable:distribution_random_fault}
		\centering
	\begin{tabular}{cc|cccc}
		\multicolumn{2}{c}{} &\multicolumn{4}{c}{$s'$}\\
			&& $00$ & $01$ & $10$ & $11$ \\
		\cline{2-6}
		\multirow{4}{*}{$s$} 
		&$00$	& $\textcolor{tumblue}{\frac{1}{4}}$ & $\frac{1}{4}$ & $\frac{1}{4}$ & $\frac{1}{4}$ \\
		&$01$	& $\frac{1}{4}$ & $\textcolor{tumblue}{\frac{1}{4}}$ & $\frac{1}{4}$ & $\frac{1}{4}$ \\
		&$10$	& $\frac{1}{4}$ & $\frac{1}{4}$ & $\textcolor{tumblue}{\frac{1}{4}}$ & $\frac{1}{4}$ \\
		&$11$	& $\frac{1}{4}$ & $\frac{1}{4}$ & $\frac{1}{4}$ & \textcolor{tumblue}{$\frac{1}{4}$} \\
	\end{tabular}
	\end{subtable}
\begin{subtable}{.24\textwidth}
	\caption{Bit-flip}
	\label{stable:distribution_bit_flip}
	\centering
	\begin{tabular}{cc|cccc}
		\multicolumn{2}{c}{} &\multicolumn{4}{c}{$s'$}\\
		&& $00$ & $01$ & $10$ & $11$ \\
		\cline{2-6}
		\multirow{4}{*}{$s$} 
		&$00$	& \textcolor{tumblue}{$0$} & $0$ & $0$ & $1$ \\
		&$01$	& $0$ & \textcolor{tumblue}{$0$} & $1$ & $0$ \\
		&$10$	& $0$ & $1$ & \textcolor{tumblue}{$0$} & $0$ \\
		&$11$	& $1$ & $0$ & $0$ & \textcolor{tumblue}{$0$} \\
	\end{tabular}
\end{subtable}
	\end{center}
	\caption{FDTs of 2-bit variables}
	\label{table:sifa_fault_distribution_table}
\end{table}

\noindent \textbf{Statistical model:}
By injecting a fault in an operation an intermediate value of $n$-bit is affected.
The intermediate value is represented by the two random variables $S$ and $S'$, before and after the injected fault.
This means the intermediate value is denoted by the random variable $S$ when no faults are present and $S'$ otherwise.
Both random variable can take values $s\in\mathcal{S} =  \{0,...,2^{n} -1\}$.
The probabilities of the individual entries of the \gls{FDT} are calculated as shown in 
\cref{eq:sifa_fdt_prob_distribution}, where $s$ corresponds to the values in the rows and $s'$ to the values in the columns of \cref{table:sifa_fault_distribution_table}.
\begin{equation}
p_s(s') := P(S' = s'~|~S = s).
\label{eq:sifa_fdt_prob_distribution}
\end{equation}
The elements of the diagonal of an \gls{FDT} correspond to the probabilities for different ineffective faults as shown in 	\cref{eq:sifa_fdt_prob_diagonal}.
\begin{equation}
	p_{s'}(s') := P(S' = s'~|~S = s').
	\label{eq:sifa_fdt_prob_diagonal}
\end{equation}
We assume the \gls{FDT} is not known to the attacker. 
Nevertheless, it is still possible to exploit the diagonal's deviation from the uniform distribution.
Since we assume the presence of a detection-based countermeasure, or an authenticated decryption like \textsc{Gimli}-decrypt (which inherently provides the required filtering), we will state the statistical model explicitly for this scenario.
Here, the attacker has only access to samples where the intermediate values under attack fulfill $S = S'$.
Under the assumption, that $S$ is uniformly distributed with $\textrm{P}(S=s) = 2^{-n} =\frac{1}{|\mathcal{S}|}$, the rate of ineffective faults $r_{\textrm{ineff}}$ can be calculated as shown in \cref{eq:sifa_ineffectivity_rate}.
\begin{equation}
	r_{\textrm{ineff}} = P( S' = S) = \sum_{s'~\in~\mathcal{S}} \frac{p_{s'}(s')}{|\mathcal{S}|}
	\label{eq:sifa_ineffectivity_rate}
\end{equation}
The diagonal of the \gls{FDT} can be expressed as conditional distribution as shown in \cref{eq_sifa_conditional_distribution_diagonal}. This distribution is later estimated by an attacker, as neither the diagonal nor $S'$  can be observed.
\begin{equation}
	p_{\textrm{ineff}} (s') = P(S' = s'~|~S' = S) = \frac{p_{s'}(s')}{|\mathcal{S}| \cdot r_{\textrm{ineff}}}
	\label{eq_sifa_conditional_distribution_diagonal}
\end{equation}
However, it is possible to calculate the hypothetical distribution $p_H$ of $S'_H$ under the assumption of a fixed key hypothesis $k_H$. 
By using the correct key guess $k_H$ the correct distribution $p_{\textrm{ineff}}(s{'}) = p_{H=\textrm{correct}}(s{'})$ is observed.
In order to distinguish it from incorrect key guesses, we assume that all distributions of incorrect key hypotheses $k_{H=\textrm{wrong}}$ are close to the uniform distribution i.e. $p_{H=\textrm{wrong}}(s') \approx \theta(s') = 2^{-n}$.
The distinguisher $D(p_H)$ is used to rank the key candidates according to their distance to $\theta(s')$. %
The chi-squared ($\chi^{2}$) test as introduced by Pearson \cite{Pearson1900}, is used to calculate a metric for the difference of two probability distribution function $a$ and $b$, both with values $x\in \mathcal{X}$ and $N$ samples as shown in \cref{eq:sifa_general_pearson}
\begin{equation}
	\chi^2(a,b) =  N \sum_{x \in \mathcal{X}} \frac{a(x) - b(x)}{b(x)}.
	\label{eq:sifa_general_pearson}
\end{equation}
Since incorrect key guesses follow the uniform distribution, we can use the $\chi^2$ metric to distinguish distributions resulting from the correct key hypothesis $p_{H={\textrm{correct}}}$ and a uniform distribution $\theta$ caused by an incorrect hypothesis.
This leads to the $\chi^2$-distinguisher as shown in \cref{eq:sifa_chi2_decision}.
\begin{equation}
	D(p_{H_i}) = \text{CHI}(p_{H_i}) := \chi^2(p_{H_i}, \theta) 
	\label{eq:sifa_chi2_decision}
\end{equation}
Alternatively, a scaled version of CHI, the \gls{SEI} \cite{Rivain2009} as shown in \cref{eq:sifa_sei_decision}, can be used.
\begin{equation}
	\begin{aligned}
	D(p_{H_i}) = \text{SEI}(p_{H_i})%
	& \hphantom{:}=  \frac{1}{|\mathcal{S}| \cdot N} \cdot \text{CHI}(p_{H_i})
	\end{aligned}
	\label{eq:sifa_sei_decision}
\end{equation}
Where $N$ denotes the number of observed decryptions under the influence of ineffective faults.

\section{Gimli}\label{sec:gimli}
\begin{figure*}[ht]
	\begin{center}
		\includegraphics[width = 0.9\textwidth]{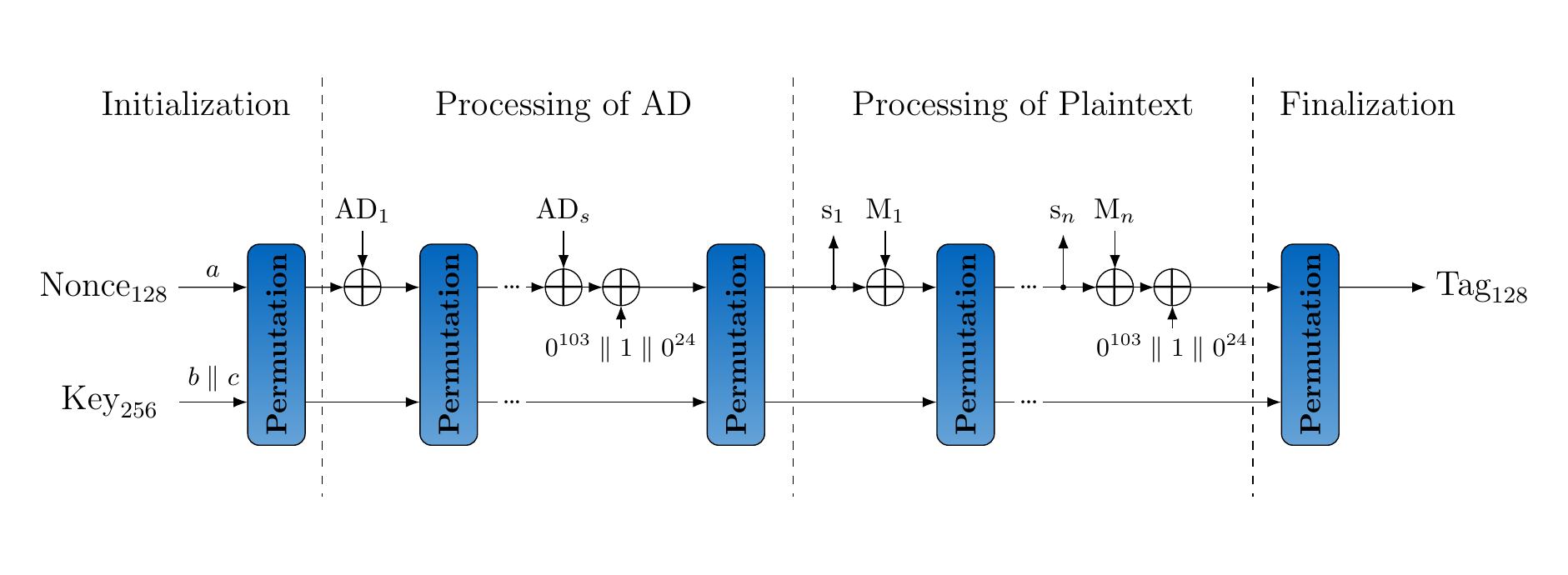}	
		\caption{Sponge construction of \textsc{Gimli}}
		\label{fig:gimli_aead_structure}
	\end{center}
\end{figure*}
\glsreset{AEAD}
\noindent \textsc{Gimli} is a suite of cryptographic primitives based on the \textsc{Gimli}-permutation by Bernstein~et~al.~\cite{Bernstein2017gimli}. 
It participates in the \gls{NIST} lightweight cryptographic project for authenticated encryption and hash.
In this paper, we focus on the \textsc{Gimli-Cipher}, a family for \gls{AEAD}. 

\subsection{The Gimli-permutation} \label{subsection:the_GIMLI_permutation}
\noindent The \textsc{Gimli}-permutation is based on an a 384-bit-state.
As shown in \cref{eq:gimli_state}, the state is defined as a $3\times4$ matrix:
the rows are named $a$,~$b$,~$c$; the columns are enumerated $0$,~$1$,~$2$,~$3$; the round is denoted by $r$.
For example $a^{11}_1$ denotes to the second $32$\,bit word before the execution of the $11$ round.
\begin{equation}
\text{State}:= 
\begin{pmatrix}
a^{r}_{0} & a^{r}_{1} & a^{r}_{2} & a^{r}_{3} \\
b^{r}_{0} & b^{r}_{1} & b^{r}_{2} & b^{r}_{3} \\
c^{r}_{0} & c^{r}_{1} & c^{r}_{2} & c^{r}_{3}
\end{pmatrix}
\label{eq:gimli_state}
\end{equation}
\vspace{0.5mm}
\noindent \cref{fig:gimli_permutation_pseudocode} describes how this state is permuted in 24 consecutive rounds. They are enumerated in reverse order: a permutation starts with round 24 and ends with round 1. During each round, the state is first substituted and permuted (\acrshort{SPBOX}). Every second round, the state is mixed linearly (alternating either a small swap or a big swap). Finally, every fourth round, a constant is added.
\begin{algorithm}\scriptsize
    \begin{algorithmic}
   	 \Function{Permute}{$a,b,c$}\Comment{Input State}
        \For{$r = 24 ~downto~ 1$}
        	\For{$j = 0 ~to~ 3$}
 			\State $\hphantom{a_0} \mathllap{t_{a}} \gets a_{j} \pmlll 24 $ \Comment{SP-Box}
			\State $\hphantom{a_0} \mathllap{t_{b}} \gets  \mathrlap{b_{j}} \hphantom{a_j} \pmlll 9 $
			\State $\hphantom{a_0} \mathllap{t_{c}} \gets \mathrlap{c_{j}} \hphantom{a_j}$
	\vspace{2mm}
			\State $\hphantom{a_0} \mathllap{a_{j}} \gets \mathrlap{t_{c}} \hphantom{t_a} \oplus t_b \oplus ((t_a~\&~t_b) \ll 3)$
			\State $\hphantom{a_0} \mathllap{b_{j}} \gets t_a \oplus t_b \oplus ((t_a~|~t_c) \ll 1)$
			\State $\hphantom{a_0} \mathllap{c_{j}} \gets t_a \oplus (t_c \ll 1) \oplus ((t_b~\&~t_c) \ll 3)$
	\vspace{0.5mm}
        	\EndFor
	\vspace{2mm}
		\If{$r~\text{mod}~4 = 0$}
			\State $a_{0}|| a_{1}|| a_{2}|| a_{3} \gets a_{1}|| a_{0}|| a_{3}|| a_{2}$\Comment{Small Swap}
	\vspace{0.5mm}
		 \ElsIf {$r~\text{mod}~4 = 2$}
			\State $a_{0}|| a_{1}|| a_{2}|| a_{3} \gets a_{2}|| a_{3}|| a_{0}|| a_{1}$\Comment{Big Swap}
	\vspace{0.5mm}
		\EndIf
	\vspace{2mm}		
		\If{$r~\text{mod}~4 = 0$}
			\State $a_{0} \gets a_{0} \oplus \texttt{0x9e377900} \oplus r$\Comment{Constant Addition}
	\vspace{0.5mm}
		\EndIf
	\vspace{0.5mm}
        \EndFor
	\State
		\Return{$(a,b,c)$}\Comment{Output State}
	\EndFunction
    \end{algorithmic}
    \caption{The \textsc{Gimli}-permutation}
    \label{fig:gimli_permutation_pseudocode}
\end{algorithm}

\subsection{The Gimli-AEAD} \label{subsection:the_GIMLI_AEAD}
\noindent The \textsc{Gimli-Cipher} is a sponge based \gls{AEAD} scheme with a \SI{128}{\bit} rate and a \SI{256}{\bit} capacity.
The rate matches $a$, the capacity matches the concatenation of $b$ and $c$.
\cref{fig:gimli_aead_structure} depicts the four phase during an \gls{AEAD}.
As the exploited fault is ineffective in respect to the output,
the exact behavior of the \gls{AEAD} scheme is secondary and a brief description of the four phases is sufficient:
First, the state is initialized with a \SI{128}{\bit} nonce, and the \SI{256}{\bit} key as depicted in \cref{eq:gimli_init}.
\begin{equation}
\begin{aligned}
 &\text{Nonce:}  & a^{24}_0 ...\, a^{24}_3                            &\leftarrow n_0\, n_1\, n_2\, n_3  \\
 &\text{Key:}  \hspace{2mm} &            b^{24}_0 ...\, b^{24}_3 || c^{24}_0 ...\, c^{24}_3 &\leftarrow k_0\, k_1~ ...~\,k_7 
\end{aligned}
\label{eq:gimli_init}
\end{equation}
Second, the associated data $AD_i$ are absorbed in blocks of \SI{128}{\bit}.
Subsequently, the message block key~$s_i$ is squeezed. Depending on the mode, either the ciphertext $C_i = M_i \oplus s_i$ or the plaintext $M_i = C_i \oplus s_i$ is generated.
In any case, next, the plaintext $M_i$ is absorbed.
Finally, the tag is calculated.
Between all phases and absorbed blocks the \textsc{Gimli}-permutation is invoked.
Incomplete blocks are padded.
Additionally, there is a domain separation between the processing of \gls{AD}, the processing of plaintext, and finalization.
For decryption, the tag is not output, but compared to the received tag.
If both tags match, the plaintext is released, otherwise, the empty string is output.

\section{SIFA on Gimli} \label{sec:sifa_gimli}
\noindent We target the decryption of \textsc{Gimli} because \gls{AEAD} schemes only release the plaintext if the computed tag matches the original tag.
This behavior can be exploited to distinguish between effective and ineffective faults, as an effective fault results in a tag mismatch. %

\subsection{Fault Injection Location} \label{subsec:sifa_gimli_pos_fault_injection}
\begin{figure*}[ht]
	\begin{center}
	\includegraphics[width = 0.9\textwidth]{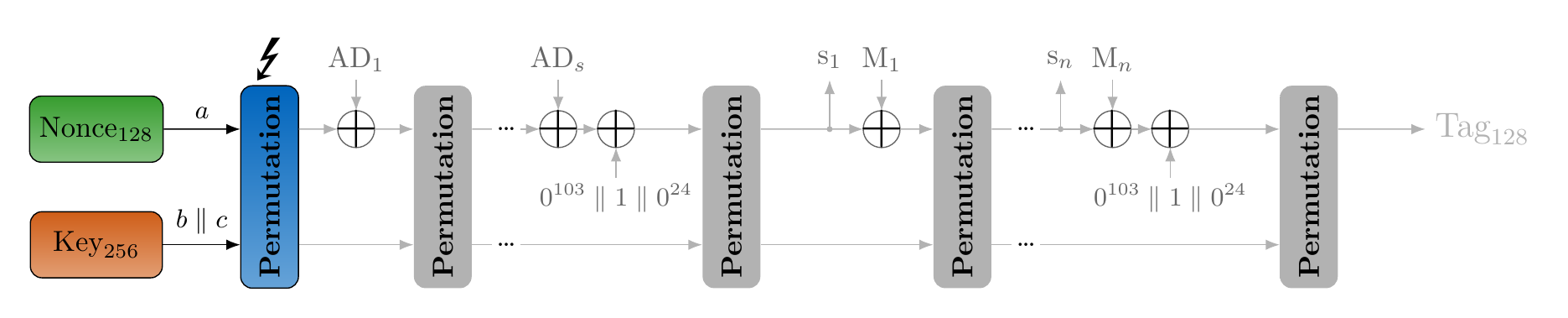}
	\caption{Fault Injection Location \textsc{Gimli}}
	\label{fig:gimli_structure_attack_point}
	\end{center}
\end{figure*}

\begin{figure}[ht]
	\includegraphics[width = 0.49\textwidth]{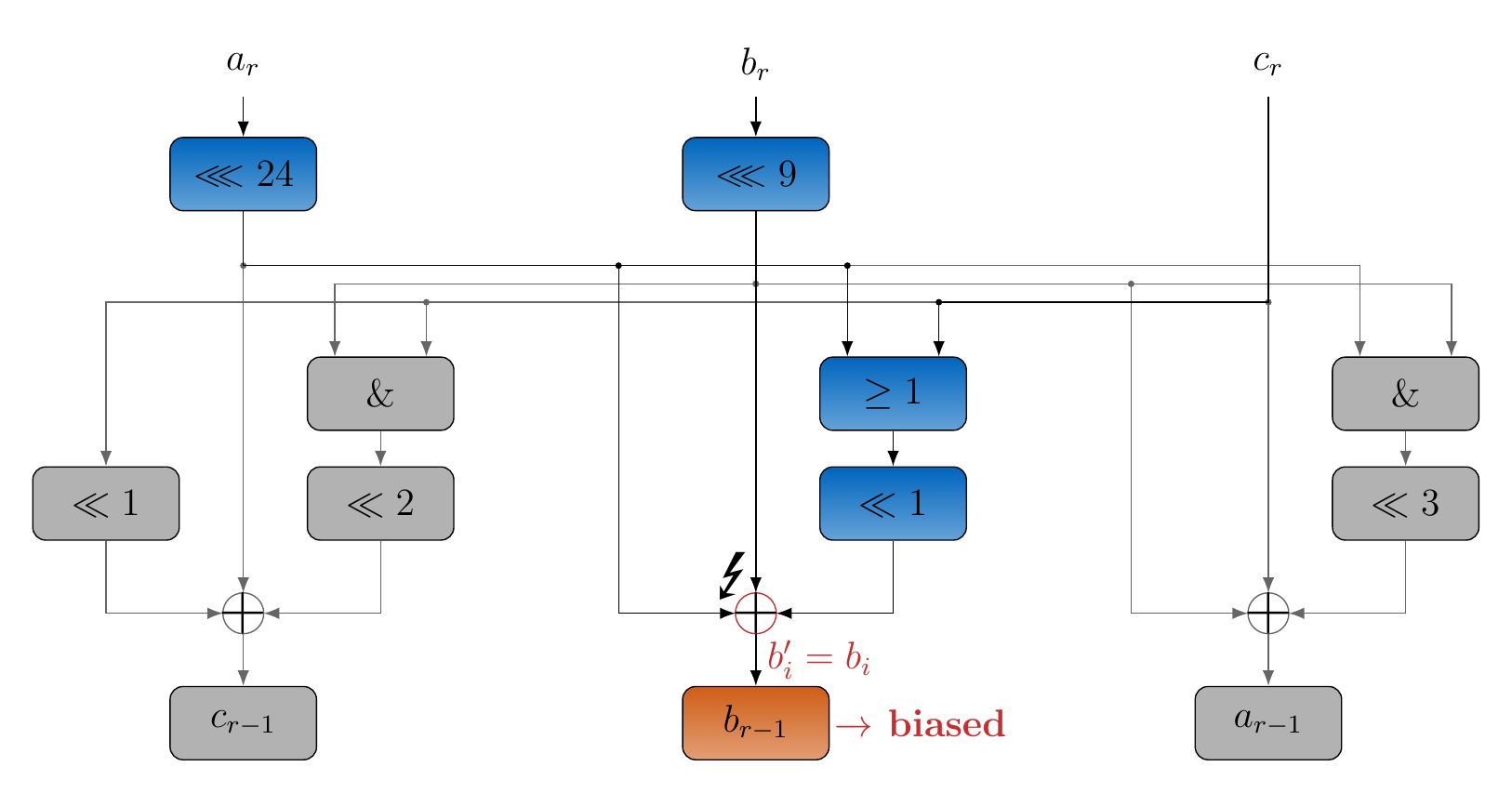}
	\caption{Fault Injection Location \gls{SPBOX}}
	\label{fig:sifa_gimli_sp_box_attack}
\end{figure}

\noindent For the \gls{SIFA} on \textsc{Gimli} we evaluated different locations where an induced ineffective fault can be exploited.
Similar to the attacks on Ketje and Keyak \cite{Dobraunig2016fault}, we use the nonce and a hypothesis of the target partial subkey $k_H$ to calculate an intermediate value of \textsc{Gimli}.
In general, this is possible for any intermediate value during the decryption of \textsc{Gimli}.
However, in order to reduce the number of involved key-bits of the %
intermediate value and the number of hypotheses $N_H$, it is desirable to attack the early rounds of the first \textsc{Gimli}-permutation.
\cref{fig:gimli_structure_attack_point} shows an attack mounted during the initialization phase. 
When we target the first \textsc{Gimli}-permutation, we can choose from one of the 24 rounds. %
The \gls{SPBOX} of the \textsc{Gimli}-permutation poses the best attack target, due to the involved non-linearity.
A possible position to inject an ineffective fault into the \gls{SPBOX} is colored \textcolor{darkred}{red} in \cref{fig:sifa_gimli_sp_box_attack}.
We mount the attack on a biased $b^r$, however a similar reasoning can be applied for a biased $a^r$ or $c^r$. 
The attacked round is a trade-off between the number of recoverable key-bits $n_\textrm{keybits}$ and number of possible key hypotheses $N_H$.
The earlier the attack, the simpler are the equations for the intermediate values, but also fewer key bits can be revealed. %
The later the attack, the higher is the number of involved key bits $n_\textrm{keybits}$ and thus, more hypotheses $N_{H}$ must be checked. 
The number of hypotheses $N_{H}$ grows exponentially with the number of involved key bits $n_\textrm{keybits}$ as shown in \cref{eq:sifa_gimli_number_hypo}.
\begin{equation}
	N_H \sim 2^{n_\textrm{keybits}}
	\label{eq:sifa_gimli_number_hypo}
\end{equation}
In order to determine the exact number of involved key bits $b_{k}$ the dependencies of an intermediate value must be traced back to the initialization phase where the state gets initialized using the known nonce and the unknown key.
\subsection{Calculation of Intermediate Values} \label{subsec:sifa_gimli_calc_intermediate}
\begin{figure}[ht]
	\centering
	\includegraphics[width = 0.49\textwidth]{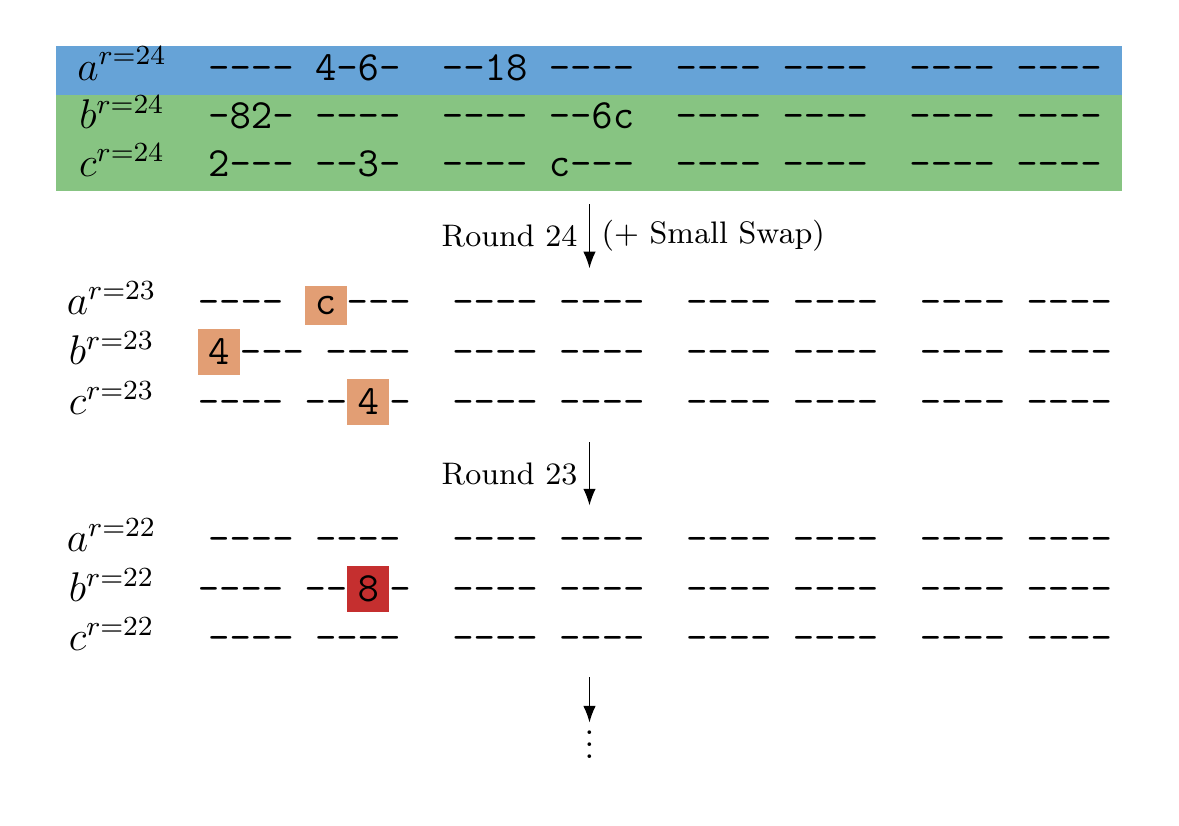}
	\caption{Dependencies of $b^{r=22}_{0,7}$}
	\label{fig:sifa_gimli_dependency_22}
\end{figure}

\begin{figure}[ht]
	\centering
	\includegraphics[width = 0.49\textwidth]{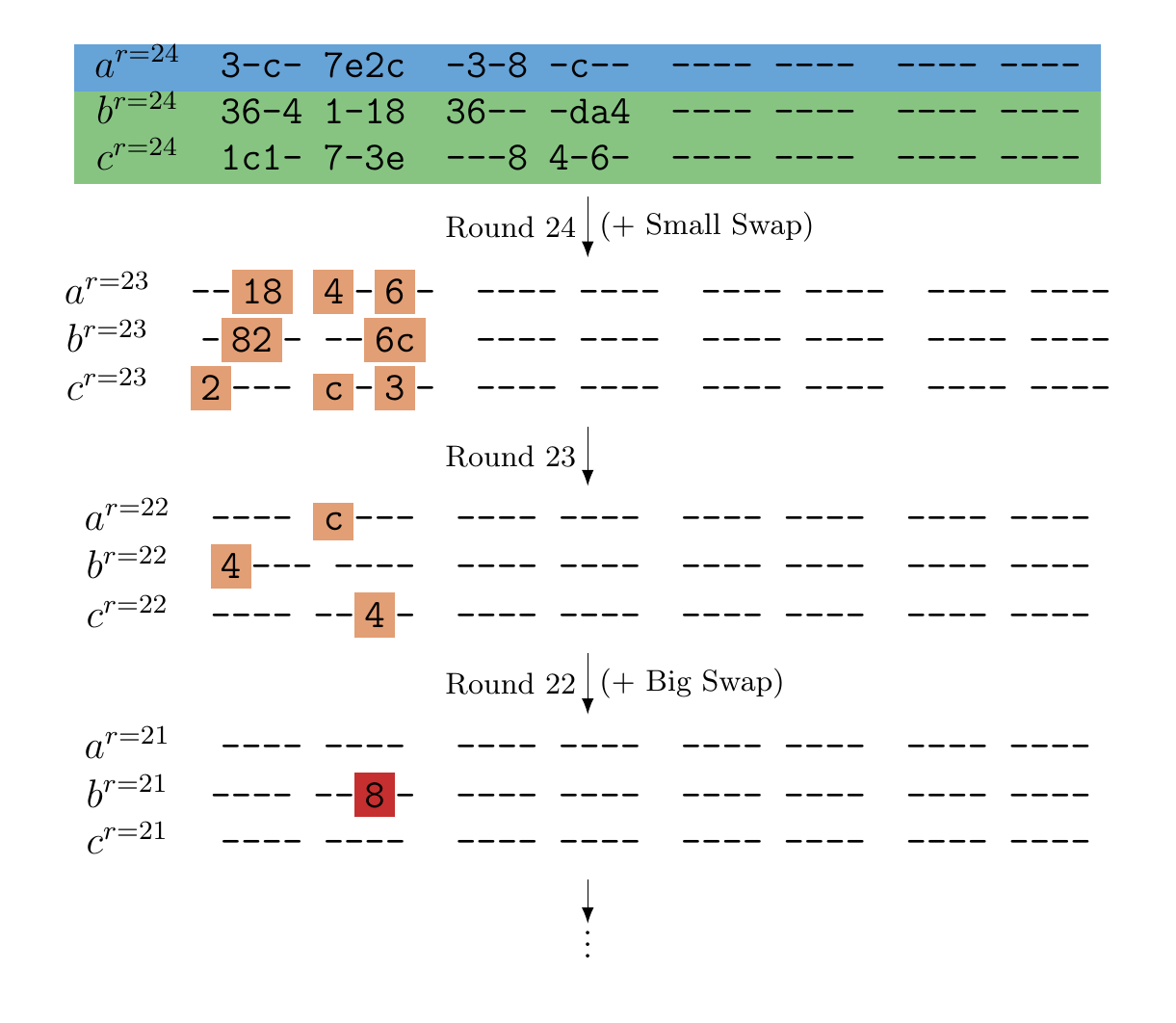}
	\caption{Dependencies of $b^{r=21}_{0,7}$}
	\label{fig:sifa_gimli_dependency_21}
\end{figure}

\noindent The dependencies of an intermediate value under attack are related to the fault injection location.
We will demonstrate an attack of bit $b^{r}_{0,7}$.
Hereby  $b^{r}_{0,7}$ denotes the seventh bit of the word $b_{0}^{22}$  during round~$r$.
The resulting dependencies of $b^{r}_{0,7}$ with respect to the according injection location are shown in the second row of \cref{tab:sifagimli_b07dependency}. %
\begin{table}[ht]
	\centering
\begin{tabular}{c c c}
	$r$  & $n_\textrm{keybits}$ &Dependencies of $b^{r}_{0,7}$ \\
\hline
	23 &2& $k_{0,31} \oplus n_{0,15} \oplus (n_{0,14} ~|~ k_{4,6})$ \\
\hline
	   && $k_{0,21} \oplus n_{0,6} \oplus (n_{0,5} ~|~ k_{4,29}) \oplus k_{5,15} \oplus k_{1,6} \oplus$\\
	22 &11& $\oplus (n_{1,20} ~\&~ k_{1,3}) \oplus c_{15} \oplus [(k_{5,14} \oplus k_{1,5} \oplus (n_{1,19} ~\&$ \\ 
	   && $\&~ k_{1,2})\oplus c_{14}) ~|~ (n_{0,14} \oplus k_{4,5} \oplus (k_{4,4} ~\&~ k_{0,27}))]$\\
\hline
	21 &37& cf. Appendix\\
\hline
\end{tabular}
\caption{Dependencies of $b^{r}_{0,7}$ for different injection locations}
\label{tab:sifagimli_b07dependency}
\end{table}
A fault injection in the very first round, i.e. after round 24, to attack the bit $b^{23}_{0,7}$ only affects two key bits and therefore, does not offer a big advantage in terms of recoverable key bit this is shown in the first row of \cref{tab:sifagimli_b07dependency}.
If the fault is injected one round later i.e. after round 23, eleven key bits are involved in the computation of the bit $b^{22}_{0,7}$ this is shown in the second row of \cref{tab:sifagimli_b07dependency}.
By biasing the intermediate bit $b^{21}_{0,7}$ again a round later, an attacker can utilize 37 involved key-bits due to the sheer length of the involved equations we stated the full dependency equation in the Appendix.
Involved key-bits lead to a dependency due to the path along which they properagate through the \textsc{GIMLI}-permutation.
The bit wise dependencies after each \textsc{GIMLI}-round are visualized similarly to Dobraunig et al.~\cite{Dobraunig2015}.
Involved bits i.e. dependencies are represented by a $1$ independent bits are represented by '0' or '-' e.g. \texttt{c=1100} means that only bit 3 and 2 of this nibble are involved in a computation.
The position of the key is colored in \textcolor{darkgreen}{green} and the nonce in \textcolor{tumblue}{blue}.
\cref{fig:sifa_gimli_dependency_22} shows, which bits are involved in the computation of the intermediate bit $b^{22}_{0,7}$ which is colored in \textcolor{darkred}{red}.
Even though 11 bits of the key are involved in the calculation of the intermediate value $b^{22}_{0,7}$, not all of them can be identified distinctively due to linear dependencies of the involved key bits.
\cref{eq:sifa_gimli_sum23} shows these linear dependencies. 
Each of the three sums are affected by three different key bits. Therefore, only the sums $k_{s1}$, $k_{s2}$ and $k_{s3}$, but not the individual key bits can be recovered.
\begin{equation}
\begin{aligned}
    k_{s1} &= \hphantom{ k_{0,21}} \mathllap{k_{0,21}}  \oplus k_{5,15} \oplus k_{1,6} \\
 	k_{s2} &= \hphantom{ k_{0,21}} \mathllap{k_{5,14}} \oplus k_{1,5}\\
	k_{s3} &= \hphantom{ k_{0,21}} \mathllap{k_{4,5}~} \oplus (k_{4,4}~\&~ k_{0,27})
\end{aligned}
	\label{eq:sifa_gimli_sum23}
\end{equation}
Thus, an attack on the intermediate value  $b^{22}_{0,7}$ reveals only the key bits $k_{4,29}$, $k_{1,3}$ and $k_{1,2}$.
However, the key sums can also be used to build hypotheses.
This results in an advantage of $2^6$ compared to brute-forcing each individual bit of the involved key bits.
An attack on the intermediate state $b^{21}_{0,7}$ already involves 37 key bits.
Taking linear dependencies into account, the number of hypotheses is $2^{22}$.
A graphical representation of the dependencies of $b^{21}_{0,7}$ is shown in \cref{fig:sifa_gimli_dependency_21}%
Going one round further ($b^{r=20}_{0,7}$) increase the number of involved key bits to 168.
However, testing $2^{168}$ hypotheses is not feasible in a reasonable amount of time.
Thus, the attack on the intermediate states in rounds 22 and 21 offer a reasonable trade-off between the number of hypotheses and recoverable key bits.
\subsection{Fault Model}
\begin{figure}[ht]
	\centering
	\includegraphics[width = 0.43\textwidth]{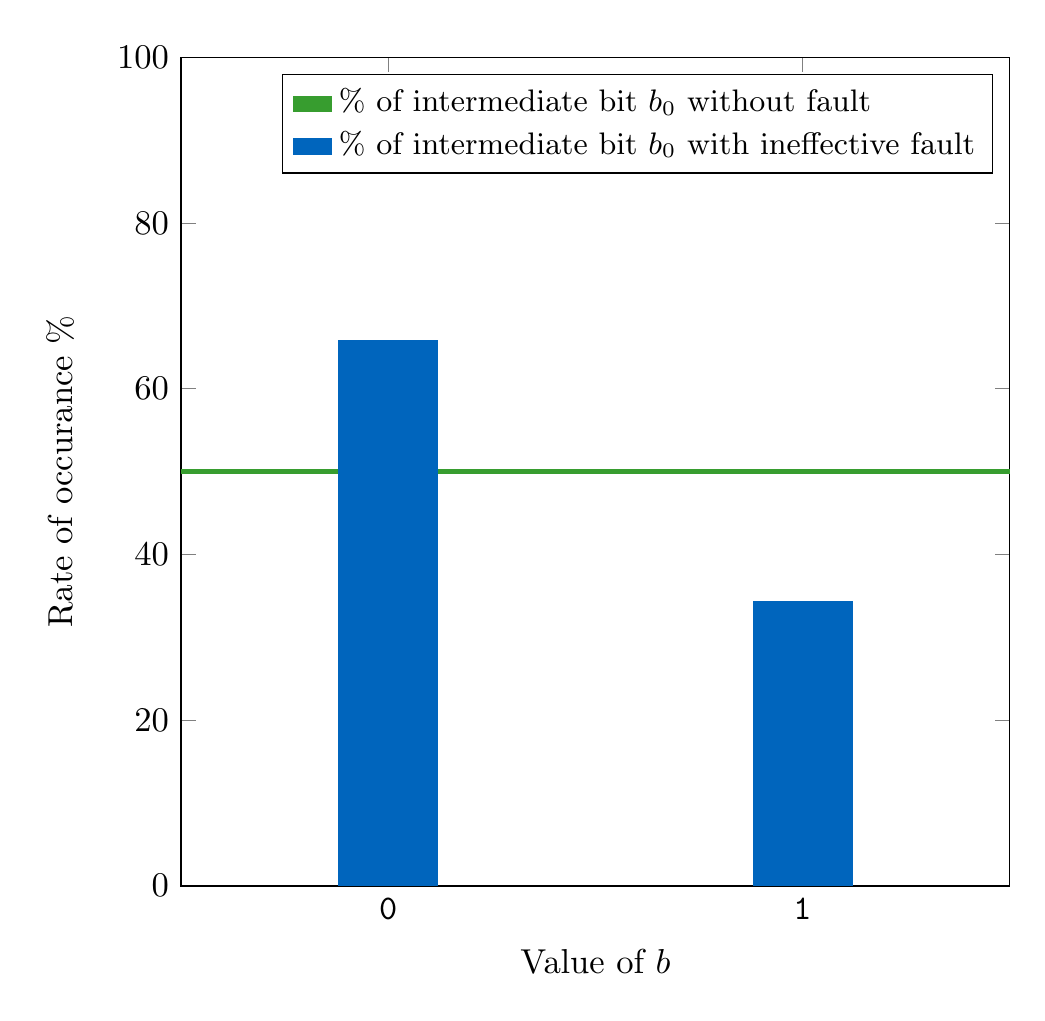}
	\caption{Histogram of intermediate values $b$}
	\label{fig:histogram_1bit_intermediate}
\end{figure}

\begin{figure}[ht]
	\centering
	\includegraphics[width = 0.43\textwidth]{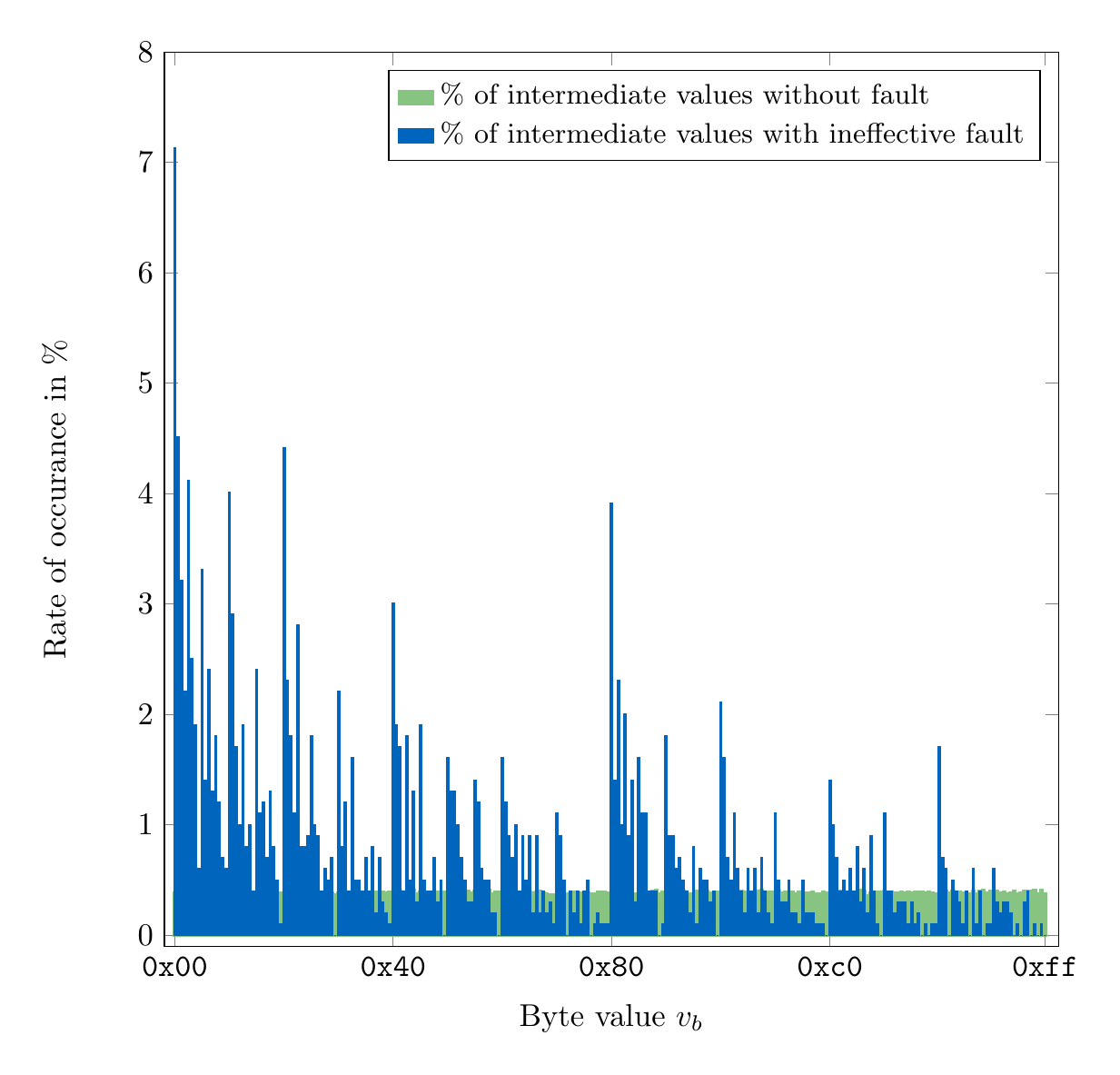}
	\caption{Histogram of intermediate values $b^{r=22}_{0,0-7}$}
	\label{fig:histogram_8bit_intermediate}
\end{figure}

\noindent In \cref{sec:sifa} the influence of some typical fault models onto the \gls{FDT}'s was shown in \cref{table:sifa_fault_distribution_table} at the example of a two bit intermediate state.
However, the fault models are not limited to 2-bit but can be applied to words with variable width $w$.
Since we cannot choose the word-width of the implementation of \textsc{GIMLI} but still want to evaluate the distribution of a single bit, it is important to evaluate, if a byte based fault model also biases each bit separately.
For example a fault of width $w=8$ is the equivalent to a byte based fault model.
We simulated faults with $w=8$ according to the probabilistic bit flip fault model where a flip from $1 \rightarrow 0$ occurs with probability $P_{1 \rightarrow 0} = \frac{2}{3}$  and a flip $0 \rightarrow 1$  with probability $P_{0 \rightarrow 1} =\frac{1}{3}$.
This biased bit flip probabilities for a one bit intermediate value $b$ result in the histogram shown in \cref{fig:histogram_1bit_intermediate}
This behavior is the same as the \gls{FDT} shown in \cref{stable:distribution_probabilistic_bitflip} which depicts the two dimensional case.
The bias of the 8-bit intermediate value $b^{r=22}_{0,0-7}$ caused by an ineffective fault is shown in \cref{fig:histogram_8bit_intermediate}.
The nearly normal distributed values without any fault are colored in \textcolor{darkgreen}{green} whereas all values leading to ineffective faults are colored in \textcolor{tumblue}{blue}.
If one compares the histogram as shown in \cref{fig:histogram_8bit_intermediate} with the previously introduced FDTs as shown in \cref{table:sifa_fault_distribution_table}, it becomes clear that this distribution can be attacked due to the deviation from the uniform distribution which is directly recognizable.
Based on the simulations as shown in \cref{fig:histogram_8bit_intermediate} we decided to use a byte based fault model i.e. $w=8$ during the explanation of the attack strategy.
\subsection{Attack Strategy} \label{subsec:sifa_gimli_attack_strategy}
\noindent For the attack it is necessary to generate decryptions under the influence of an ineffective fault.
As a result of \textsc{Gimli} being a \gls{AEAD} scheme the collected decryptions are all under the influence of an ineffective fault otherwise there would be no output due to a tag mismatch.
After a sufficient number of decryptions $N_{d}$ is obtained
we calculate the hypothetical intermediate bit $inter$.
The calculation of the intermediate bit is done with respect to the possible key hypotheses and all obtained nonces $n$.
The distribution of $inter$ is then ranked by according to the \gls{SEI}.
Now that each hypothetical distribution has been assigned an \gls{SEI}, the correct key hypothesis can be identified as the one with the largest \gls{SEI}.
The algorithmic representation of the attack is shown in \cref{fig:sifa_gimli_attack_pseudocode}.
First the hypothetical intermediate values are calculated for all possible keys.
Then for each key hypothesis the \gls{SEI} of the intermediate distribution is calculated and stored.
If a new \gls{SEI} is greater than or equal to the old one the corresponding key hypothesis is used as the new correct hypothesis.
After all hypotheses have been processed, the algorithm terminates.
\begin{algorithm}[ht]\scriptsize
    \begin{algorithmic}
	    \State $N_H \gets 2^{n_\textrm{keybits}}$
	    \State $n[N_d] \gets \texttt{loadNonces}()$
	    \State maxSEI $\gets 0$
	    \State corrHypo $\gets 0$
        \For{$i = 0 ~to~N_H$}
        	\For{$j = 0 ~to~N_d$}
			\State $ inter[i][j] \gets \texttt{calcIntermediateBit}(i,n[j])$
		\EndFor
		\State $  $
		\State $\textrm{SEI}[i] \gets \texttt{calcSEI}(inter[i][*])$
		\State $  $
		\If{$\textrm{SEI}[i] \geq \textrm{maxSEI}$}
			\State maxSEI $\gets \textrm{SEI}[i]$
			\State corrHypo $\gets i$
		\EndIf
        \EndFor
	\State
	    \Return{corrHypo}
    \end{algorithmic}
    \caption{SIFA on \textsc{Gimli}}
	\label{fig:sifa_gimli_attack_pseudocode}
\end{algorithm}
In our attack strategy, it is necessary to choose an appropriate intermediate value to attack as the involved key bits which can be recovered for each intermediate value differ.
As introduced in \cref{subsec:sifa_gimli_calc_intermediate}, it is possible to calculate some key bits directly and some key bits only in the form of a sum.
The computation of the intermediate state $b^{22}_{0,7}$ involves 11 key bits as shown in the dependency equations in \cref{tab:sifagimli_b07dependency}.
However, 8 key bits influence $b^{22}_{0,7}$ only in the form of a sum as shown in \cref{eq:sifa_gimli_sum23}.
The computation of $b^{r=22}_{0,7}$ involves $n_\textrm{keybits}=11$ but only hypotheses on 6 key bits are required as the remaining key bits only appear in the form of a sum.
Therefore, the number of hypotheses $N_{H}$ shrinks from $2^{11}$ to $2^6$.
The same effect also occurs when we target the intermediate value at the same position one round later i.e. $b^{r=21}_{0,7}$.
The computation of the intermediate state $b^{21}_{0,7}$ involves 37 key bits as shown in the dependency equations in \cref{tab:sifagimli_b07dependency}.
The key-bits $k_{4,26}$, $k_{4,20}$, $k_{4,5}$, $k_{4,4}$, $k_{4,3}$, $k_{4,2}$, $k_{4,1}$, $k_{1,26}$, $k_{1,25}$, $k_{1,8}$, $k_{1,7}$, $k_{1,2}$, $k_{0,26}$, $k_{0,25}$ and $k_{0,18}$ can be determined uniquely.
Furthermore, the sum of 22 key bits in the form of 7 sum can be determined.
From the 37 involved key bits we are able to obtain an advantage of 22 bits compared to the brute force effort over all involved key bits.
Due to some ambiguity in the large equation for $b^{r=21}_{0,7}$ the \gls{SIFA} reveals three candidates with the same \gls{SEI} after 340 decryptions under the influence of ineffective faults.
The ambiguity is caused by some nonce bits that do not differ when they cause an ineffective fault.
Although the described ambiguity is present, there is always the correct key-hypothesis among those three candidates.
By the injection of 8-bit ineffective fault, hypotheses can be build on 8 intermediate bits that can be evaluated simultaneously with almost no extra computational effort.
With this we get an advantage of at most $8\cdot6=48$ bits when attacking round 22 and at most $8\cdot 22=176$~bits when attacking round 21.
In order to obtain the complete key which is loaded during the initialization phase of \textsc{Gimli} it is necessary to repeat the proposed attacks with varying intermediate states under attack until all key bits are recovered. %
\section{Results}\label{sec:results}
\noindent Now that we have clarified the prerequisites for the attack, we will present the results.
First we will evaluate the influence of the fault width $w$ on the ineffectiveness rate of the injected faults.
Second the obtained results for the attack on $b^{r=22}_{0,7}$ and $b^{r=21}_{0,7}$ are presented.
Both attacks exploit the bias of an ineffective fault with fault width $w=8~\textrm{bit}$ injected after round 23 respectively 22 under the assumption of a probabilistic bit flip fault model.
\subsection{Influence of fault width on ineffectiveness rate} \label{subsec:res_ineffectiveness_rate}
\begin{figure}[ht]
\centering
\includegraphics[width = 0.43\textwidth]{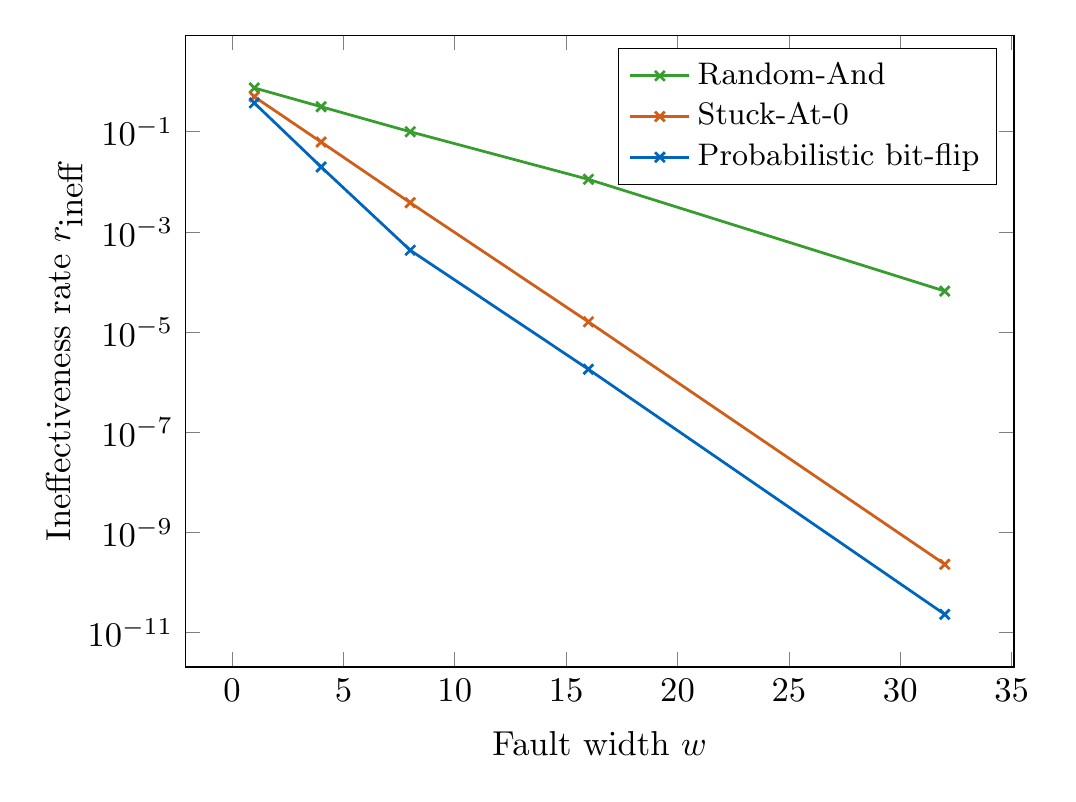}
\caption{Ineffectiveness rate~$r_{\textrm{ineff}}$ of different fault  models}
\label{fig:res_ineff_rate}
\end{figure}

\noindent \gls{SIFA} exploits the bias present in an intermediate state independently of the assumed fault model.
In practice, it is usually assumed that an attacker has no information about the \gls{FDT} which is caused by the ineffective fault injection.
Nevertheless, the only prerequisite for a successful attack is that an intermediate value follows a biased distribution.
The biased distribution is indicated by the diagonal of the \gls{FDT} which follows a non-uniform distribution as shown in \cref{table:sifa_fault_distribution_table}.
Furthermore, we typically cannot choose the target architecture where \textsc{Gimli}-\gls{AEAD} is run on, which can either be a software implementation running on a microcontroller or a hardware implementation running on a FPGA or ASIC.
If we consider the case of a software implementation of \textsc{Gimli}-\gls{AEAD}, then the fault width $w$ will be the same as the word with of the micro-controller.
If we consider a hardware implementation of \textsc{Gimli}-\gls{AEAD} the fault width $w$ is usually dependent of the implementation.
In the following we use the typical fault models random-And, Stuck-at-0 and probabilistic bit-flip exemplary to simulate a fault on a software implementation, the same reasoning can also be applied to hardware implementations.
Faults are injected during round 23 of the first \textsc{Gimli}-permutation on state $b^{r=22}_{0}$.
The width $w$ of the fault ranges from 1 to 32 bit with $w\in \{1,4,8,16,32\}$.
By calculating the number of ineffective faults $n_{\textrm{ineff}}$ divided by the number of total encryptions $N$ we obtain the ineffectiveness rate as shown in \cref{eg:ineff_rate}.
\begin{equation}
	r_{\textrm{ineff}} = \frac{n_{\textrm{ineff}}}{N}.
	\label{eg:ineff_rate}
\end{equation}
The ineffectiveness rate $r_{\textrm{ineff}}$ with respect to the fault width $w$ is shown in \cref{fig:res_ineff_rate}.
As one can see the ineffectiveness rate $r_{\textrm{ineff}}$ decreases almost linearly with the assumed fault width $w$.
The linear decrease of the ineffectiveness rate occurs independently of the three fault models.
The ineffectiveness rate of the probabilistic bit flip fault model is dependent on the assumed bit flip probabilities we decided to use this model to provide a worst case estimation.
In practice this means that attacking a 32-bit software implementation of \textsc{Gimli}, should be feasible according to  \cref{fig:res_ineff_rate}.
Especially the Random-And model offers a significant rate of ineffective faults at $w=32~\textrm{bit}$.
However, due to the very low ineffectiveness rate for the other fault models, a number of more than $10^9$ total encryptions is required for the attack. 
For the sake of simplicity further simulations where done with a fault width $w=8~\textrm{bit}$ in order to minimize the computational effort of generating ineffective faults.

\subsection{Attack on $b^{r=22}_{0,7}$}
\begin{figure}[ht]
	\centering
	\includegraphics[width = 0.49\textwidth]{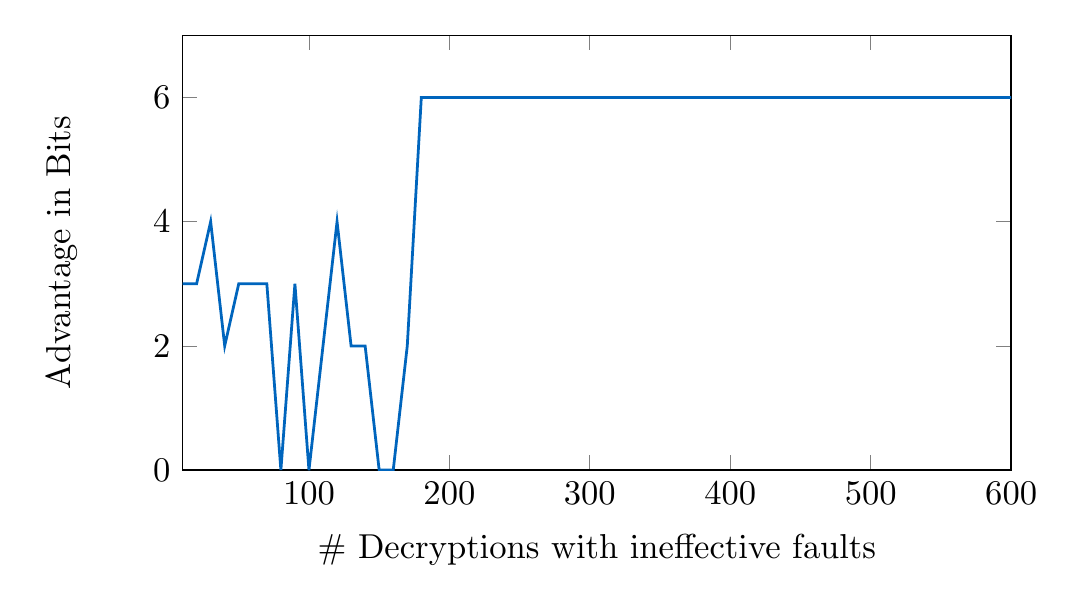}
	\caption{Advantage - Attack on round 22}
	\label{fig:res_advantage_r22}
\end{figure}

\begin{figure}[ht]
	\centering
	\includegraphics[width = 0.49\textwidth]{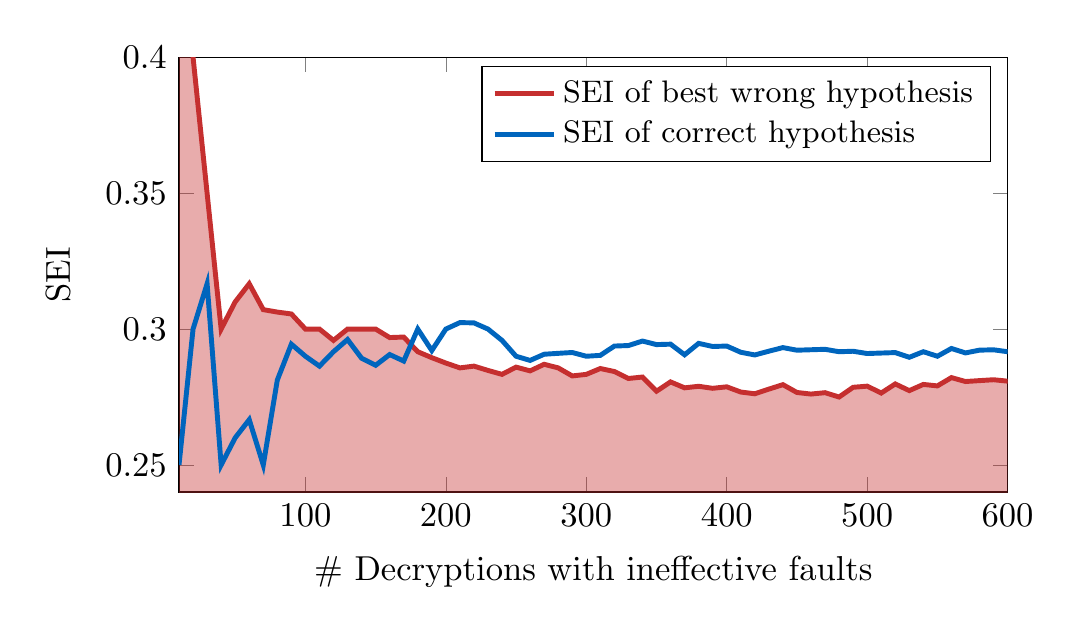}
	\caption{\gls{SEI} - Hypotheses, round 22}
	\label{fig:res_sei_r22}
\end{figure}

\noindent The attack on the intermediate state $b^{r=22}_{0,7}$ is able to retrieve the involved key bits correctly after approximately 180 ineffective faults. 
The number of required encryptions under the influence of an ineffective fault is shown in \cref{fig:res_sei_r22} where we used the  \gls{SEI} as statistical metric.
In \cref{fig:res_sei_r22} the best wrong hpyothesis is colored \textcolor{red}{red} and the correct hypothesis in  \textcolor{tumblue}{blue}.
Furthermore, it is important to notice, that after the point both are crossing line, the correct hypothesis keeps a significantly higher value.
\cref{fig:res_advantage_r22} shows the advantage over brute forcing when increasing the number of decryptions with ineffective faults.
The maximum advantage is defined as the number of unique definable parameters when attacking the single bit $b^{r=22}_{0,7}$, i.e. the three key-bits and the three sum-values therefore the maximal advantage of the attack on round 21 can be \SI{6}{\bit}.
The unstable advantage at the beginning is caused by multiple key hypotheses with similar \gls{SEI} values,
which leads to frequent change of the key hypothesis having the current maximum \gls{SEI}.
Although the correct key hypothesis is retrieved after 180 ineffective faults, some bits of a wrong hypothesis still are equal to the corresponding bits in the correct guess leading to an advantage of less than 6 bits.

\subsection{Attack on $b^{r=21}_{0,7}$}
\begin{figure}[ht]
	\centering
	\includegraphics[width = 0.49\textwidth]{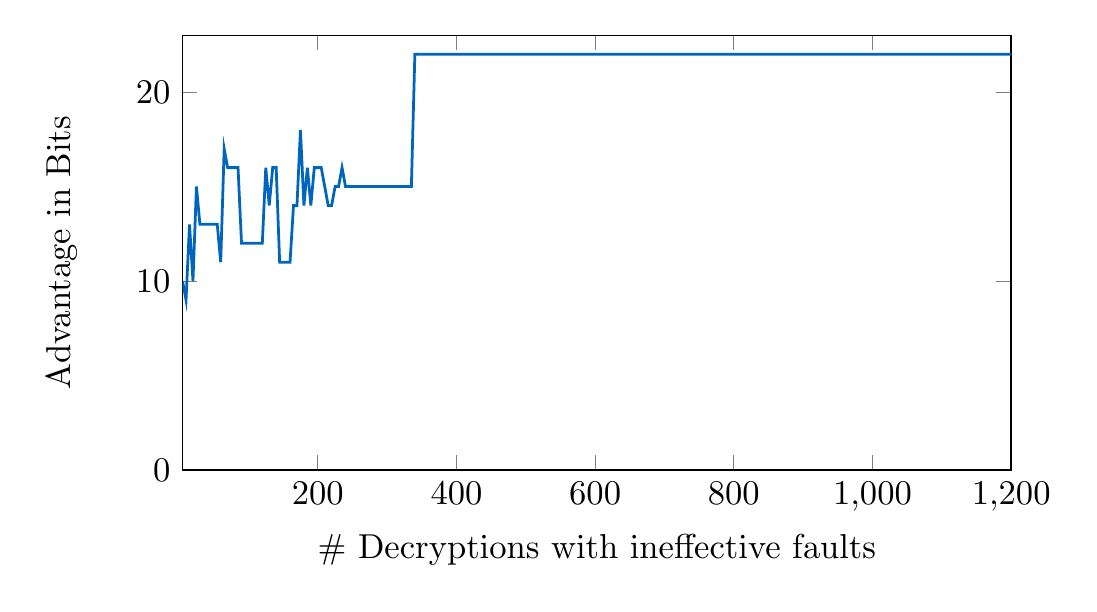}
	\caption{Advantage - Attack on round 21}
	\label{fig:res_advantage_r21}
\end{figure}

\begin{figure}[ht]
	\centering
	\includegraphics[width = 0.49\textwidth]{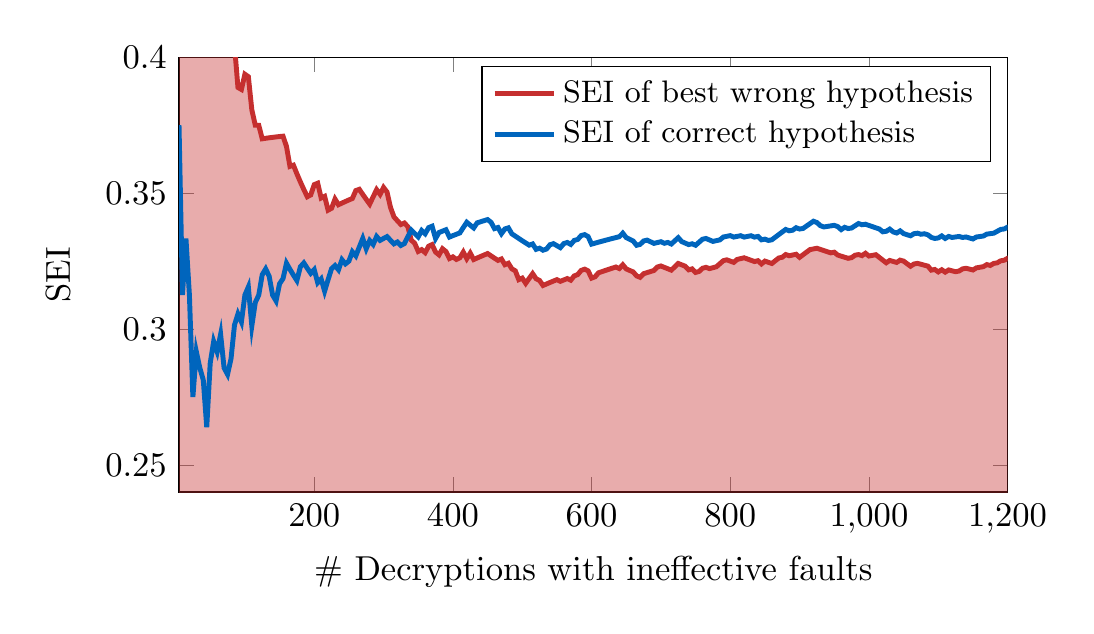}
	\caption{\gls{SEI} - Hypotheses, round 21}
	\label{fig:res_sei_r21}
\end{figure}

\noindent The attack on the intermediate state $b^{r=21}_{0,7}$ is able to retrieve the involved key bits correctly after approximately 340 ineffective faults. 
Again, the number of required encryptions under the influence of an ineffective fault is shown in \cref{fig:res_sei_r21} where we used the  \gls{SEI} as statistical metric.
In \cref{fig:res_sei_r21} the best wrong hpyothesis is colored \textcolor{red}{red} and the correct hypothesis in  \textcolor{tumblue}{blue}.
\cref{fig:res_advantage_r21} shows the advantage over brute forcing when increasing the number of decryptions with ineffective faults.
The possible advantage when attacking $b^{r=21}_{0,7}$ is 22-bits at max.
Although the hypothesis with highest \gls{SEI} changes frequently when using less than 340 ineffective faults, the correct key-guess has the maximal \gls{SEI} after obtaining it. 
\section{Conclusion}\label{sec:conclusion}
\noindent In this work we presented the \gls{SIFA} of the \gls{AEAD} scheme \textsc{Gimli} (\textsc{Gimli-24-Cipher}).
Furthermore, we investigated the influence of the fault width $w$ on the rate of ineffective faults $r_{\textrm{ineff}}$.
The fact that \gls{SIFA} can be applied to the \gls{AEAD} scheme \textsc{GIMLI} should be considered a threat.
This is mainly due to the fact that the fault model assumed by \gls{SIFA} is rather simple to achieve i.e. a biased intermediate value which is processed during a cryptographic operation.
Due to the ineffective characteristic common countermeasures against fault attacks can be circumvented by \gls{SIFA}.
\section*{Acknowledgements} 
\ack{
\noindent We would like to thank the anonymous reviewers for their valuable comments and suggestions on the paper, as these helped us to improve it.
This work was partly funded by the German Federal Ministry of Education and Research in the project HQS through grant number 16KIS0616.} \printbibliography
\section*{Appendix}  \label{sec:appendix}

\subsection{Dependency $b^{r=21}_{0,7}$ - round 22} \label{appendix:b21_r22}
\begin{equation*}
b^{r=21}_{0,7} = b^{r=22}_{0,30} \oplus a^{r=22}_{0,15} \oplus (a^{r=22}_{0,14} ~|~ c^{r=22}_{0,6}).
\end{equation*}

\subsection{Dependency $b^{r=21}_{0,7}$ - round 23} \label{appendix:b21_r23}
\begin{equation*}
\begin{aligned}
b^{r=22}_{0,30}&= b^{r=23}_{0,21} \oplus a^{r=23}_{0,15} \oplus (a^{r=23}_{0,5}~|~ c^{r=23}_{0,29})\\
a^{r=22}_{0,15}&= c^{r=23}_{0,15} \oplus b^{r=23}_{0,6} \oplus (a^{r=23}_{0,20}~\&~b^{r=23}_{0,3})\\
a^{r=22}_{0,14}&= c^{r=23}_{0,14} \oplus b^{r=23}_{0,5} \oplus (a^{r=23}_{0,19}~\&~b^{r=23}_{0,2})\\
c^{r=22}_{0,6} &= a^{r=23}_{0,14} \oplus c^{r=23}_{0,5} \oplus (c^{r=23}_{0,4} ~\&~b^{r=23}_{0,27})\\
\end{aligned}
\end{equation*}

\subsection{Dependency $b^{r=21}_{0,7}$ - input} \label{appendix:b21_r24}
\begin{equation*}
\begin{aligned}
	b^{r=23}_{0,21}&= \textcolor{tumblue}{k_{0,12}} \oplus n_{0,29} \oplus (n_{0,28}~|~ k_{4,20})\\
	a^{r=23}_{0,6} &= \textcolor{tumblue}{k_{5,6}} \oplus \textcolor{tumblue}{k_{1,29}} \oplus (n_{1,11} ~\&~k_{1,26}) \oplus c_{6}\\
	a^{r=23}_{0,5} &= \textcolor{tumblue}{k_{5,5}} \oplus \textcolor{tumblue}{k_{1,28}} \oplus (n_{1,10} ~\&~k_{1,25}) \oplus c_{5}\\
	c^{r=23}_{0,29}&= n_{0,5} \oplus \textcolor{tumblue}{k_{4,28}} \oplus (\textcolor{tumblue}{k_{4,27}} ~\&~\textcolor{tumblue}{k_{0,18}})\\
\end{aligned}
\end{equation*}
\vspace{0.2cm}

\begin{equation*}
\begin{aligned}
	c^{r=23}_{0,15}&= n_{0,23} \oplus \textcolor{tumblue}{k_{4,14}} \oplus (\textcolor{tumblue}{k_{4,13}}~\&~\textcolor{tumblue}{k_{0,4}})\\
	b^{r=23}_{0,06}&= \textcolor{tumblue}{k_{0,29}} \oplus n_{0,14} \oplus (n_{0,13}~| ~k_{4,5})\\
	a^{r=23}_{0,20}&= \textcolor{tumblue}{k_{5,20}} \oplus \textcolor{tumblue}{k_{1,11}} \oplus (n_{1,25}~\&~k_{1,8}) \oplus c_{20}\\
b^{r=23}_{0,03}&= k_{0,26} \oplus n_{0,11} \oplus (n_{0,10}~| ~k_{4,2})\\
\end{aligned}
\end{equation*}
\vspace{0.2cm}

\begin{equation*}
\begin{aligned}
	c^{r=23}_{0,14}&= n_{0,22} \oplus \textcolor{tumblue}{k_{4,13}} \oplus (\textcolor{tumblue}{k_{4,12}}~\&~\textcolor{tumblue}{k_{0,3}})\\
	b^{r=23}_{0,05}&= \textcolor{tumblue}{k_{0,28}} \oplus n_{0,13} \oplus (n_{0,12}~| ~k_{4,4})\\
	a^{r=23}_{0,19}&= \textcolor{tumblue}{k_{5,19}} \oplus \textcolor{tumblue}{k_{1,10}} \oplus (n_{1,24}~\&~k_{1,7}) \oplus c_{19}\\
b^{r=23}_{0,2} &= k_{0,25} \oplus n_{0,10} \oplus (n_{0,9} ~| ~k_{4,1})\\
\end{aligned}
\end{equation*}
\vspace{0.2cm}

\begin{equation*}
\begin{aligned}
	a^{r=23}_{0,14}&= \textcolor{tumblue}{k_{5,14}} \oplus \textcolor{tumblue}{k_{1,5}} \oplus (n_{1,19}~\&~k_{1,2}) \oplus c_{14};\\
c^{r=23}_{0,05}&= n_{0,13} \oplus k_{4,4} \oplus (k_{4,3} ~\&~k_{0,26})\\
c^{r=23}_{0,04}&= n_{0,12} \oplus k_{4,3} \oplus (k_{4,2} ~\&~k_{0,25})\\
b^{r=23}_{0,27}&= k_{0,18} \oplus n_{0,3} \oplus (n_{0,2} ~| ~k_{4,26})\\
\end{aligned}
\end{equation*}

\subsection{$b^{r=21}_{0,7}$ - key sums} \label{appendix:b21_sums}
\noindent  Key bits which can only recovered in the form of a sum are colored in  \textcolor{tumblue}{blue}.

\end{document}